\pdfoutput=1

\documentclass[preprint]{sig-alternate-05-2015}

\usepackage[utf8]{inputenc}
\usepackage[T1]{fontenc}
\usepackage[english]{babel}
\usepackage{enumitem}
\usepackage[table]{xcolor}
\usepackage{graphics}
\usepackage[hang,footnotesize]{subfigure}
\newcommand\autoref[1]{Figure~\ref{#1}}
\usepackage[protrusion,expansion,tracking,kerning,spacing]{microtype}
\microtypecontext{spacing=nonfrench}
\SetTracking{encoding={*}, shape=sc}{0}
\usepackage{balance}

\newcommand\tr[2]{T(#1)/\discretionary{}{}{}R(#2)}

\begin{document}

\clubpenalty=10000
\widowpenalty=10000

\title{Preference Between Allocentric and Egocentric 3D~Manipulation in a Locally Coupled Configuration}
\subtitle{\vskip -1.25em} 
%
%
%
%
%

\numberofauthors{5} 
%
\author{
%
%
\alignauthor Paul Issartel\\
       \affaddr{LIMSI-CNRS}\\
       \affaddr{Orsay, France}\\
       \email{paul.issartel@limsi.fr}
\alignauthor Lonni Besançon\\
       \affaddr{INRIA}\\
       \affaddr{Saclay, France}\\
       \email{lonni.besancon@inria.fr}
\alignauthor Florimond Guéniat\\
       \affaddr{LIMSI-CNRS}\\
       \affaddr{Orsay, France}\\
       \email{contact@gueniat.fr}
\and  
\alignauthor Tobias Isenberg\\
       \affaddr{INRIA}\\
       \affaddr{Saclay, France}\\
       \email{tobias.isenberg@inria.fr}
\alignauthor Mehdi Ammi\\
       \affaddr{LIMSI-CNRS}\\
       \affaddr{Orsay, France}\\
       \email{mehdi.ammi@limsi.fr}
}

\maketitle
\begin{abstract}
We study user preference between allocentric and egocentric 3D~manipulation on mobile devices, in a configuration where the motion of the device is applied to an object displayed on the device itself. We first evaluate this preference for translations and for rotations alone, then for full 6-DOF manipulation. We also investigate the role of contextual cues by performing this experiment in different 3D~scenes. Finally, we look at the specific influence of each manipulation axis. Our results provide guidelines to help interface designers select an appropriate default mapping in this locally coupled configuration.
\end{abstract}

%
%
%




\section{Introduction}
With the rising availability of mobile devices and their increasing processing power, 3D~applications are becoming more common on these devices. A fundamental part of interaction with such software is 3D~manipulation~\cite{bowman04}, i.e.~translations and rotations in 3D~space. While most mobile devices rely on touch-based control, tactile input requires a 2D to 3D mapping for 3D~manipulation. Several projects~\cite{issartel16,neale13} thus proposed to use the built-in motion sensors found in many mobile devices to provide full 3D~input, by detecting the device's own motion and mapping it to virtual objects. Recently, the Tango tablet\footnotemark{} provided a major technological step forward by combining gyroscopes, accelerometers, and visible/infrared cameras to fully track its translations and rotations relative to the surrounding environment. This mode of interaction is thus likely to become more widely used in the future.

\begin{figure}[t]
  \centering
  \includegraphics[width=\linewidth]{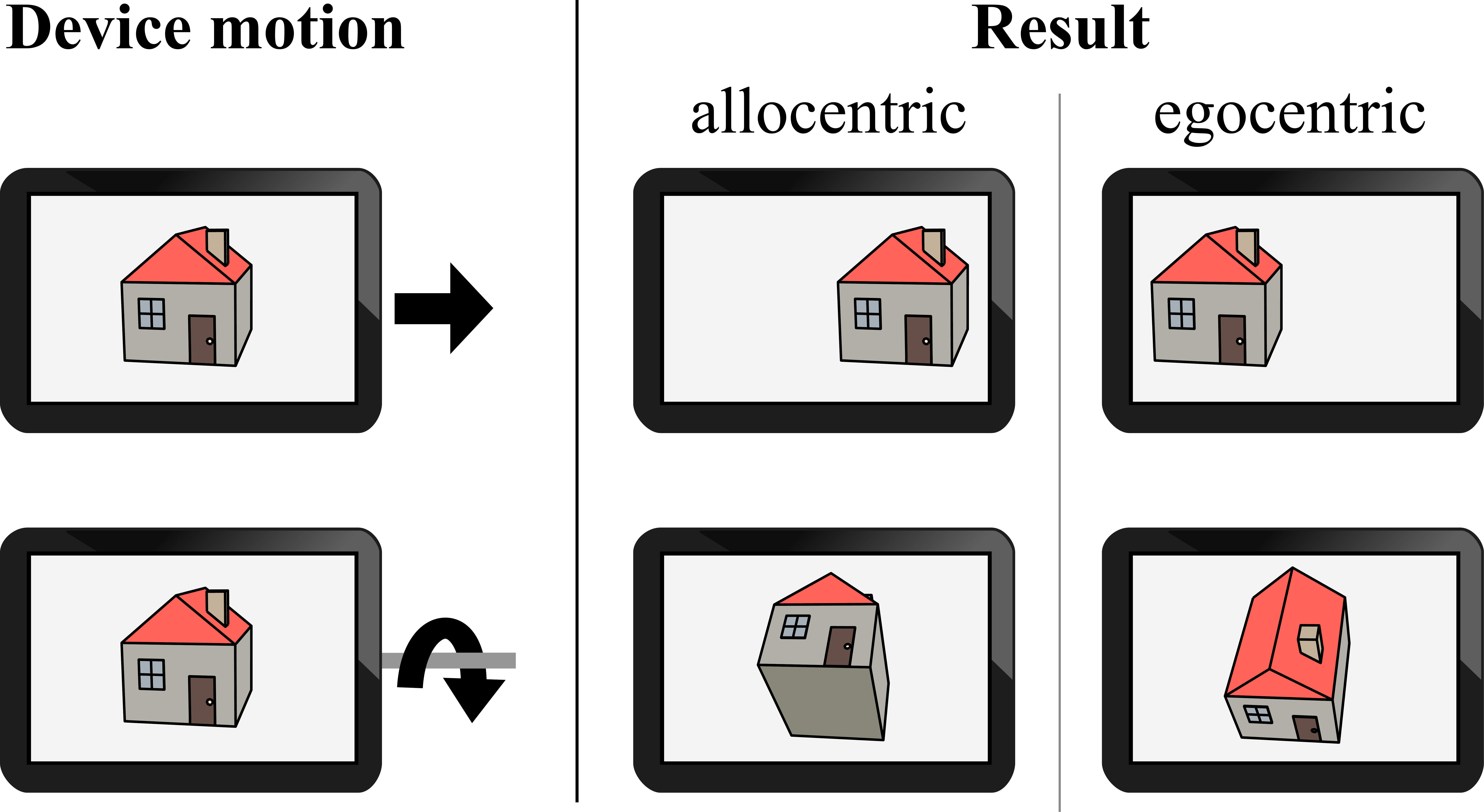}
  \caption{Illustration of the allocentric and egocentric mappings.}
  \label{fig:allo-ego}
\end{figure}

A unique aspect of mobile devices is that they integrate input and display capabilities in the same device. This means that users are holding both the input device and the display device in their hands---a ``locally~coupled'' configuration. For this reason, using the motion of a mobile device for 3D~input can be interpreted in two ways (see \autoref{fig:allo-ego}):
\begin{itemize}
\item the mobile device could be seen as a ``handle'' to control 3D~objects---the \emph{allocentric} interpretation---or
\item the mobile device could be seen as a ``window'' that moves around 3D~objects---the \emph{egocentric} interpretation.
\end{itemize}
In the allocentric interpretation, the manipulated 3D~objects move in the \emph{same} direction as the ``handle'' represented by the mobile device. In the egocentric interpretation, objects move in the \mbox{\emph{opposite}} direction, as if the viewpoint itself was controlled through a handheld~``window''.
\looseness=-1

Although either alternative can be obtained by simply reversing the direction of motion in the control-display mapping~\cite{issartel16}, the question remains of which mapping should be actually implemented. This is not only relevant if the interface does not offer a way to switch between the two alternatives, but it also affects first-time users. The concept of ``compatibility''~\cite{chan03}, originating from the ergonomics literature, states that a chosen mapping should correspond to the alternative \emph{most often expected} among the population. Choosing a compatible mapping is thus essential for good usability and minimal fatigue~\cite{chan03}, and to avoid errors such as accidental inversions~\cite{diaz05}. Even after extensive training, evidence suggests that a non-compatible mapping still results in reduced user performance~\cite{chan03}.

In this paper we thus examine which mapping users prefer in various situations, in order to help interface designers choose which mapping to implement in each~case.
\footnotetext{\url{http://www.google.com/atap/project-tango/}}

\section{Related work}

\subsection{Allocentric and egocentric reference frames}

Each interpretation of the locally coupled configuration is a matter of the cognitive relationship between the objects in space, called a \emph{reference frame}.
The literature on spatial cognition generally distinguishes between two fundamental reference frames~\cite{klatzky98}: allocentric (also called exocentric) and egocentric. However, most of this literature focuses on the relationship between the user and his or her surrounding objects. In our case, there is both a relation between the user and the mobile device, and a relation between the mobile device and the 3D~objects displayed on~it. We thus need to examine how we can apply these terms to this specific~situation.
\looseness=-1

Klatzky~\cite{klatzky98} defines the two terms in the context of whole-body navigation: egocentric as being related to the ``perspective of the perceiver,'' and allocentric as being related to an external, independent framework. Burgess~et~al.~\cite{burgess04} studied these reference frames in a spatial updating task (judging the relative motion of objects between an initial and a final configuration), which has a certain degree of similarity with our situation. They define the allocentric frame as ``the association of object locations to external landmarks'' and the egocentric frame as being related to ``self-motion.'' Poupyrev~et~al.~\cite{poupyrev98} define these reference frames in the context of immersive environments: egocentric interaction techniques are those linked to the avatar's viewpoint, while ``exocentric'' (i.e.~allocentric) techniques are those performed from an external location. In a study by Diaz and~Sims~\cite{diaz05} on accidental inversions, the egocentric condition was viewed from the operators' eyes and the allocentric condition was viewed from outside their body.

Overall, the egocentric term seems to be associated with the idea of the viewing perspective, and the allocentric term with the idea of a fixed, external reference point. In our case, when the manipulated object moves in the same direction as the mobile device, the object appears to be directly controlled by the device's motion and thus to move relative to the surrounding space. We~therefore describe such mappings as ``allocentric.'' When the manipulated object moves in the opposite direction, the mobile device appears to directly control the \emph{perspective} on the object. We~therefore call such mappings~``egocentric.''

\subsection{Population stereotypes}

The relationship between an input device's motion and the motion's result on a display has long been investigated in \mbox{ergonomics}. Although this relationship---or ``mapping''---can take many forms, some of them better match the user's mental model. Such mappings are said to be \emph{compatible}~\cite{fitts53,chan03}. An important goal for interactive systems design is thus to determine which mapping is most compatible with the target population.

When faced with several functionally equivalent alternatives such as the two mappings we study, the option which is most often expected among the population is called a \emph{population stereotype}~\cite{fitts51,wiebe09}. Several population stereotypes have been identified in previous research. Warrick's principle~\cite{warrick47,wiebe09}, for instance, states that the controlled display should move in the same direction as the side of the input device that is closest to it. The clockwise-to-increase principle~\cite{wiebe09} states that the controlled display should ``increase'' or ``move up'' when the input device is rotated clockwise. These principles, however, were established under the assumption that the input device is separated from the display. In our locally-coupled configuration, in contrast, they are both the same object: the mobile device. Existing population stereotypes thus cannot be applied unless they have been validated under a locally-coupled setting.
\looseness=-1

\subsection{Preferred mapping in a locally-coupled\\ configuration}

Many interfaces have been proposed that use the motion of a mobile device to control objects on the device's screen. Only few of these works, however, explicitly mention the existence of two alternative mappings (direct/allocentric and inverted/egocentric) and the rationale behind the final choice.

Rekimoto~\cite{rekimoto96} proposed a menu interface controlled by device tilt. Both mappings were discussed, but the chosen alternative (moving the menu behind a fixed cursor, equivalent to our egocentric interpretation) was selected for technical reasons rather than based on an user study. Weberg~et~al.~\cite{weberg01} chose the opposite option (moving the cursor in a fixed menu, equivalent to our allocentric interpretation) in their tilt-based menu interface, on the basis that it ``felt very intuitive and natural.'' Bartlett~\cite{bartlett00} mentioned the existence of two groups of users with different mental models, each expecting the controlled picture to move in an opposite direction when tilting the device. Hinckley and~Song~\cite{hinckley11} also mentioned that slightly more than half of their users had an opposite mental model to others in their tilt-to-zoom technique. 
Although all this work was conducted in a locally-coupled configuration, it still does not provide sufficient evidence in favor of either the allocentric or the egocentric interpretation. In addition, the studies cited above were conducted on~1D or 2D~interfaces, and may thus not be generalizable to \mbox{3D~manipulation}.

\subsection{Preferred mapping for 3D~manipulation}

Kaminaka and~Egli~\cite{kaminaka85} investigated the preferred mapping to translate and rotate a cube through a lever. The lever was alternatively mapped to translations or rotations along each axis. Although this is an actual 3D~manipulation task, the 1D~input device and non locally-coupled configuration make the results of this study difficult to generalize to our case. Diaz and~Sims~\cite{diaz05} investigated ``accidental inversions'' of rotations, i.e.~what happens when users encounter a mapping opposite to their expectation. Such inversions allow one to identify the actual population stereotypes for 3D~rotations. However, the study used a 2D~mouse as input device and an external display, which again makes the results difficult to apply to our case.

There appears to be a single study that is fully applicable to our case: Issartel~et~al.~\cite{issartel16} studied the preferred mapping for 3D~manipulation tasks on a locally-coupled mobile device. This work revealed some marked stereotypes in the studied population. However, the study was only preliminary and the number of participants~(10) was relatively low for producing reliable results. Even though both translation and rotation mappings were considered, full 6-DOF mappings were not. Finally, the use of an external tracking marker in the environment could have created a bias toward the egocentric interpretation, a limitation mentioned in the study itself. We thus use Issartel~et~al.'s~\cite{issartel16} work as a basis but greatly expand the experimental protocol, apparatus, and number of participants to produce broader, deeper, and more reliable results.

\section{Measuring user expectation}

It is challenging to determine the ``expected'' choice of allocentric or egocentric mapping: many experimental biases can affect a study such as prior exposure to the interface, learning effects, or even how the interaction is described to users. We considered several ways to determine this expected mapping.
\looseness=-1

\begin{figure*}
  \includegraphics{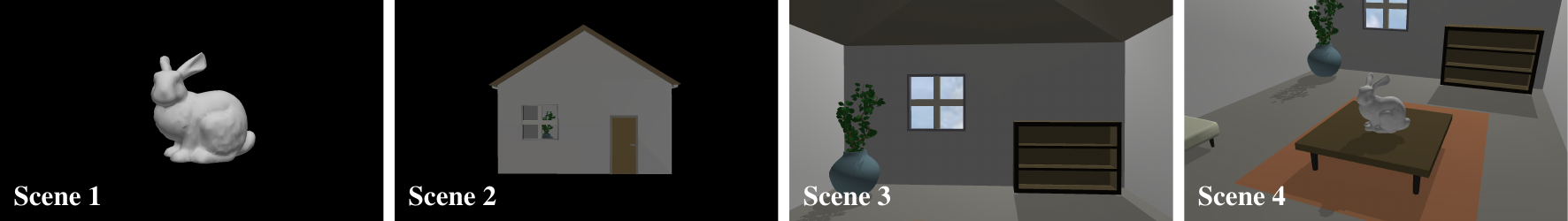}
    \caption{The four scenes used in the experiment. In Scene~3, the house model is controlled in the same way as in Scene~2 but the viewpoint is located inside the house. In Scene~4, only the object on the table is controlled by the user while the surrounding environment remains fixed.}
  \label{fig:scenes}
\end{figure*}

The first one, inspired by previous work on population stereotypes~\cite{wiebe09} and spatial updating~\cite{burgess04}, would be to show participants a non-interactive description of a manipulation task: for instance, an image representing an object at an initial location, and another image of the same object at a target location. Then, participants would be asked which direction they would move the mobile device to obtain the target result. This protocol has the advantage of only providing the minimum amount of information needed to answer the question, thus avoiding many confounding factors. However, it also has an important drawback: participants never actually use the interaction technique. In our case, it might be difficult for participants to answer questions about this possibly unfamiliar mode of interaction without having experienced it beforehand.

Another way would be to ask participants to perform multiple 3D manipulations in both allocentric and egocentric modes, record the resulting trajectories, and analyze them to detect accidental inversions~\cite{diaz05}. Such inversions can provide an objective indication that the mapping did not match the user's expectations. This protocol also gets the participants to actually use the interface, although there could be learning effects from prolonged use. It is challenging and error-prone, however, to reliably detect accidental inversions in full 3D~trajectories as produced by untrained users. Participants may possibly be taught how to generate ``clean'' trajectories with extensive training, but this training may also distort their preference between the two mappings compared to the general population.

A third way would be to have participants perform object manipulation tasks in both allocentric and egocentric modes, then \emph{rate} each mapping. Participants would rate a mapping depending on whether the manipulated object moves and rotates in the direction they expected (which we call the ``naturalness'' rating). Again, this protocol lets participants actually use the interface, but the actual evaluation now consists of a fully subjective assessment. Compared to the previous approach, this has the benefit of being practically feasible even with novice users, avoiding the biases associated with training. Furthermore, it is also possible that some participants may find both mappings acceptable---or reject both. By letting participants rate both mappings rather than simply choosing the ``better'' one, this protocol can provide a more detailed understanding of the actual user preference between the two mappings. Therefore, we decided to use this last protocol in our experiment.

\section{Contextual cues}

Previous work~\cite{burgess04,diaz05} has shown that the expected reference frame is not only a matter of personal interpretation, but can also be affected by cues from the environment, i.e.~\emph{contextual cues}.  In our configuration, contextual cues would come from the \emph{virtual environment} because we focus on what happens on the mobile device's screen. Since the virtual environment necessarily varies from one system to another, it is essential to determine whether and how the virtual scene visible on the screen can influence the expected~mapping.
\looseness=-1

First, we can hypothesize that the nature of the manipulated virtual object itself may have an influence. If the object looks like it would be readily manipulable in the real world (e.g.,~a figurine or a fruit), users may expect to be able to move it directly, i.e.~the allocentric interpretation. In contrast, if the object looks like it could be part of the scenery (e.g.,~a~house or a landscape model), users may expect to move around it rather than manipulate it themselves---the egocentric interpretation.

Second, there could be an influence of the geometrical relationship between objects in the virtual scene. If the manipulated object visually moves on the screen whereas other objects in the scene remain fixed (relative to the device), then the surrounding objects may be perceived as an environment relative to which the manipulated object is moving, favoring an allocentric interpretation. If the manipulated object is viewed from inside, then the object is perceived as surrounding the mobile device, which may reinforce the interpretation that the mobile device is \emph{moving inside} the object (egocentric interpretation) rather than moving the object (allocentric interpretation).

\section{Hypotheses and settings}

Based on these thoughts we had the following hypotheses about which mapping would be expected in different situations:
\begin{itemize}
\item \textbf{H1}:\,\,When the manipulated object is viewed from inside, users expect an egocentric mapping;
\item \textbf{H2}:\,\,When the manipulated object is moving within a fixed virtual environment, users expect an allocentric mapping;
\item \textbf{H3}:\,\,When the manipulated object represents a typically static part of a virtual environment (e.g.~a house), users expect an egocentric mapping.
\end{itemize}

Although Issartel et al.~\cite{issartel16} seemingly disproved hypothesis~\textbf{H2}, they suspected a bias caused by the presence of a fixed marker in the real environment, which could have led participants toward an egocentric mental model. We thus wanted to re-test \textbf{H2} in a markerless tracking setup without this bias.

In order to test the above hypotheses, we designed four different virtual scenes with different contextual cues (\autoref{fig:scenes}):
\looseness=-1
\begin{enumerate}
\setlength\itemsep{0pt}
\item a generic object (Stanford rabbit) on a black background, serving as the baseline scene;
\item an object more likely to be perceived as static (a~house), on~a black background;
\item the house seen from inside;
\item the same object as in Scene~1, surrounded by a fixed house.
\end{enumerate}

Previous studies (e.g.,~\cite{diaz05,issartel16,kaminaka85}) often considered translations and rotations separately. The use of device motion as input modality, however, allows full 6-DOF manipulation. It is thus important to also consider both components simultaneously. To investigate the role of each component in 6-DOF manipulation tasks, we stated the following null hypothesis:
\begin{itemize}
\item \textbf{H4}:\,\,Having a ``correct'' mapping (which matches user expectations) is equally important for translations and for rotations, i.e.~when performing translations and rotations simultaneously, the rating is equally affected by the choice of translation mapping as by the choice of rotation mapping.
\end{itemize}

Finally, some studies~\cite{diaz05,kaminaka85} have revealed different user expectations between the axes of manipulation. We therefore hypothesized to find such differences in our configuration:
\begin{itemize}
\item \textbf{H5}:\,\,Some axes are more important than others in the perceived naturalness of the resulting mapping.
\end{itemize}

\section{Apparatus}

The experiment was entirely self-contained in a single Tango tablet, providing display, input, and tracking capabilities. The virtual scene was displayed on the tablet's screen together with several tactile buttons to control the experiment. The tablet continuously tracked its own motion in the real world using its built-in sensors. There was thus no external marker in the environment. We captured positions and orientations relative to the fixed initial location where the software was started.

We used this tracking information to implement a relative position control mapping similar to Issartel et al.~\cite{issartel16}. We chose this mapping for its directional compliance so that participants could focus on the \emph{sense} of motion (allocentric or egocentric) without confusing it with with the \emph{axis} of motion (which always matches device motion in a directional compliant mapping). We also added a clutching mechanism: the device motion was only applied to the manipulated object while a finger touched the tablet screen. Participants could thus interrupt manipulation to reposition the tablet during complex tasks.

Participants were seated on a chair during the experiment, holding the tablet in landscape mode with both hands. The chair stood in the middle of the room as the presence of nearby fixed objects (e.g., a desk) could have biased participants toward an egocentric interpretation. We used a non-swivel chair to encourage participants to rotate the mobile device itself during rotation tasks rather than rotating themselves on the~chair.

\section{Participants}

To get more generalizable results than existing exploratory work~\cite{issartel16}, we used a larger and broader participant pool. We recruited 30~unpaid participants (12~females) whose age ranged from~20 to~53 (mean=30.3, SD=10.5). Among them, 22~had a university degree while 8~had a high-school degree or less. They all had normal or corrected-to-normal vision.

With this broader user sample it became possible to investigate an additional question: whether the familiarity with 3D~software can influence the preference between allocentric and egocentric mappings. However, only 7~participants reported to have sufficient knowledge of 3D~modeling software, which was too small to conduct such an analysis. On the other hand, 14~participants---nearly half of them---reported to regularly play 3D~video games. We thus chose to focus on gaming experience as an indicator of familiarity with 3D~software. Perhaps unsurprisingly, the group of video game players was largely correlated with younger age and male gender. Still, we assumed that the gaming experience itself would have more influence on this experiment than age or gender.

\section{Procedure and task}

We first presented participants with the tablet device and told them they would have to ``perform translations and rotations in four different virtual environments.'' In our explanations we took great care to avoid terms such as ``translating/rotating the object'' or ``moving in the virtual scene'' since any such mention could have led participants toward an allocentric or egocentric interpretation. There was no prior training phase---we wanted to avoid biases associated with previous usage of the interface. Instead, we demonstrated how to perform translations/rotations by moving and rotating the tablet in front of the participant, and demonstrated the clutching mechanism by pressing and releasing a finger on the screen. The tablet's screen was blanked during this tutorial step.

Participants were then asked to conduct the experiment without further instructions. The experiment itself consisted of a series of \emph{conditions}, in which two or more mappings were to be evaluated in a given virtual environment. On the bottom of the tablet's screen, several buttons labeled~C1,~C2, etc.\ represented the mappings, with the mappings randomly assigned to them. Pressing a button activated the corresponding mapping so participants could switch between mappings to rate them. Except for the third part of the experiment (see below), participants were free to change or go back and forth between mappings at any time during a condition.

To help them assess their own preference, participants were asked to perform 3D~manipulation tasks under each mapping. These tasks consisted in translating/rotating a 3D~object to a target location, i.e.~a docking task. In a typical docking task, the target is normally visually represented in the virtual scene. However, we could not display this target in every condition since having a fixed object in the scene would have resulted in the situation mentioned in hypothesis~\textbf{H2}. Moreover, we could not describe this task in terms of ``moving an object to a target'' as it would have biased participants toward an allocentric interpretation. We thus printed images of the target locations~(\autoref{fig:scenes}) on a physical sheet of paper, attached to the wall in front of the participants, and which they could consult any time wanted during the study. On each trial, the manipulated object started at a different position and/or orientation. Participants were asked to ``try to obtain the same result'' as in the images, by any means involving translations or rotations---thus without forcing them into an allocentric or egocentric interpretation. When the target location was reached, the manipulated object changed color to indicate success and was moved to a new location when the finger was released. Participants were encouraged to repeat this task several times to form an accurate opinion before rating a mapping.

When ready to give a rating, participants pressed a button on the tablet's screen and were presented a Likert scale ranging from ``not natural'' to ``natural''.
We explained the meaning of ``natural'' to participants as ``whether your actions produce translations/rotations in the direction you expected.'' Again, this definition was carefully worded to avoid any allocentric or egocentric formulation.

Since the Tango tracking system sometimes exhibits a small drift there was a risk that this could lead to unnatural ratings under the above definition. We thus also told participants that ``if a slight continuous motion ever occurs without any action on your part, this is a technical limitation that you should ignore in your rating.'' When all mappings in a condition were rated, the next condition was automatically started.
\looseness=-1

\subsection{Part 1: Translation and rotation}

In the first part of the experiment, translations and rotations were evaluated separately. Therefore, the conditions consisted of translation-only tasks and rotation-only tasks in the four environments, under two mappings: allocentric and~egocentric.

Scene~1---the most generic one---was always presented first to serve as a ``baseline'' with minimal learning biases. The first two conditions were thus translation tasks in Scene~1 and rotations tasks in Scene~1, presented in an alternate order between participants. The remaining conditions were the 6~combinations of translations and rotations with each of the three other scenes, presented in a random order.
Ratings were given on a 4-point Likert scale. This scale was specifically selected to lack a ``neutral'' point and to encourage participants to decide whether they perceived a mapping as natural or not.

\subsection{Part 2: Simultaneous translations/rotations}

In a second part of the experiment, translations and rotations were performed concurrently (i.e.,~full 6-DOF~manipulation) and each component was alternatively made allocentric of egocentric. There were thus 4~mappings to be evaluated: the four combinations of allocentric or egocentric translations with allocentric or egocentric rotations. These mappings were evaluated within each virtual scene, themselves presented in a random~order. We again used a 4-point Likert scale to establish comparisons with the results of the first part.

\subsection{Part 3: Per-axis inversion}

The third part was optional. Since the parts 1--2 already took approximately~30~min, we asked participants to continue voluntarily.

14~participants agreed to continue and with them we examined two conditions, one for translations and one for rotations, presented in random order. Both conditions were set up in Scene~1. In both conditions there were 8~different mappings, in which each manipulation axis ($x$,~$y$, and~$z$) was alternatively inverted.
We then asked participants to rate the techniques on a 3-point Likert scale, thus turning the rating into a choice between ``not natural,'' ``neutral,'' and ``natural''. We deliberately reduced the rating scale and allowed a neutral point to not overwhelm participants, given the large number of mappings to compare and the small changes between them.

\section{Results and discussion}

In our analysis we focus on \emph{effect sizes}---i.e.,~how much the ratings given to a mapping differ from the ratings given to another mapping---to investigate which mapping was preferred in each condition.
The ratings obtained from Likert scales are \emph{ordinal} data, so we used non-parametric statistical tests to quantify the effect sizes. For each condition we performed a Wilcoxon-Pratt signed-rank test on the ratings given to each mapping alternative. We then computed a normalized effect size~$r$ from the $z$~statistic produced by this test, as per~Fritz~et~al.~\cite{fritz12}.

Guidelines for the effect size~\cite{fritz12} are that $r{>}$0.5 is a large effect, $r{>}$0.3 is a medium effect and $r{>}$0.1 is a small effect, but these limits should not be seen as hard thresholds. We also report a bootstrapped standard error~$\sigma_r$ for each effect size.
\looseness=-1

\subsection{Translations}

\begin{figure}[t]
  \centering
  \includegraphics[width=\linewidth]{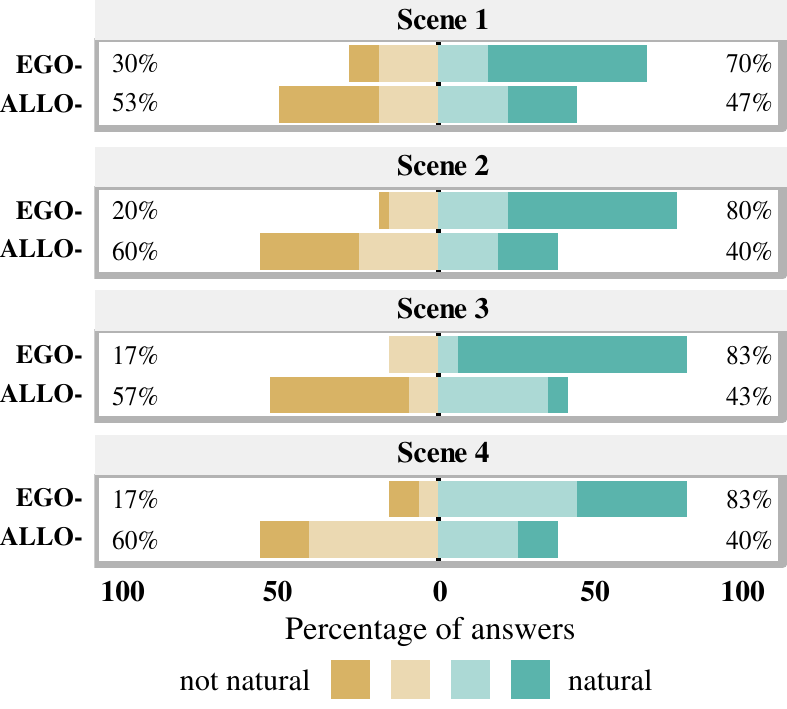}
  \caption{Participants' ratings for each mapping in the translation tasks.}\vspace{-1ex}
  \label{fig:part1-trans}
  \vspace*{-2.5mm}
\end{figure}

\autoref{fig:part1-trans} summarizes the ratings given by participants to each mapping in the translation-only tasks.

Scene~1 was always presented first in order to minimize potential biases acquired during manipulation, and was meant as a neutral environment without any of the contextual cues present in the other scenes. Therefore, the ratings obtained from this scene should best approximate the participants' \mbox{``baseline''} mental model. Although the previous experiment by Issartel~et~al.~\cite{issartel16} showed a strong preference for egocentric translations in such a neutral condition, there was a suspicion that this result could have been biased by the presence of a visible marker in the environment. Because we eliminated this marker in the present setup we had no reason anymore to expect that one mapping would be preferred over the other.
\looseness=-1

Nevertheless, translation ratings in Scene~1 revealed a small to medium preference in favor of egocentric translations ($r{=}0.27$, $\sigma_r{=}0.11$). While not as definite as in previous results~\cite{issartel16}, this preference nevertheless appears to remain true in our markerless setup. The distribution of answers shows that the egocentric mapping was indeed found natural by most participants, whereas the allocentric mapping led to mixed ratings. The egocentric mapping was thus clearly preferred by participants on this first approach to our interface, although the allocentric mapping was not completely rejected~either.

Scene~2 showed a similar pattern to Scene~1, with egocentric translations being preferred over allocentric translations. The effect size was actually higher ($r{=}0.41$, $\sigma_r{=}0.09$), though the standard error makes this distinction not completely certain. This second scene was specifically designed to test hypothesis~\textbf{H3} that a typically unmovable 3D~object would favor an egocentric mapping. A stronger preference for egocentric translations would thus tend to support hypothesis~\textbf{H3}. Yet, even if confirmed this effect appears to be quite small. In addition, since the ``baseline'' mapping for translations already seem to be egocentric (as demonstrated in Scene~1), an effect that reinforces the egocentric mapping would have little practical implications for the choice of a default translation~mapping.

In Scene~3, the preference was strongly in favor of the egocentric mapping ($r{=}0.53$, $\sigma_r{=}0.05$), even more than in Scenes~1 and~2. There is thus strong evidence to support hypothesis~\textbf{H1}, i.e., that translating an object viewed from inside is preferably accomplished with an egocentric mapping.

Scene~4 also showed an egocentric mapping preference ($r{=}0.30$, $\sigma_r{=}0.12$). This is surprising since we were expecting that manipulating an object within a fixed virtual environment would favor an allocentric mental model (hypothesis~\textbf{H2}). When we noticed during the experiment that some participants gave unexpectedly high ratings to the egocentric mapping, we took the opportunity to ask them the reasons behind this choice at the end of the first session. Their comments suggested that they were mainly focused on performing the task and did not pay much attention to the fixed virtual scene. Indeed, since participants went through at least two other conditions before encountering Scene~4, it is believable that the task was beginning to become a ``routine'' at this point. In addition, since the scene was fixed in screen space it is plausible that some participants merely considered it as a background image and did not adopt the mental model that we expected them to do.

\subsection{Rotations}

\begin{figure}[t]
  \centering
  \includegraphics[width=\linewidth]{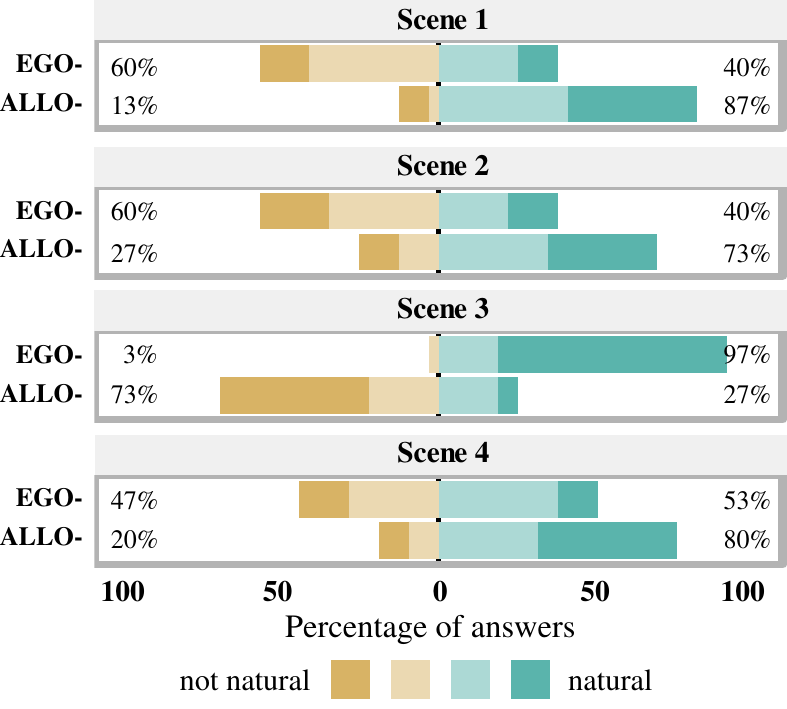}\vspace{-.5ex}
  \caption{Participants's ratings for each mapping in the rotation tasks.}
  \label{fig:part1-rot}
  \vspace*{-2.5mm}
\end{figure}

\autoref{fig:part1-rot} summarizes participants' ratings for the rotation-only tasks.
The ``baseline'' condition Scene~1 showed a medium effect ($r{=}0.37$, $\sigma_r{=}0.10$) in favor of allocentric rotations. Most participants rated the allocentric mapping as natural but gave mixed ratings to the egocentric mapping. This confirms the results of previous work~\cite{issartel16} which also showed a preference for allocentric rotations in this condition.

The allocentric mapping was also preferred in Scene~2, but apparently less strongly ($r{=}0.24$, $\sigma_r{=}0.12$) than in Scene~1. Although the standard error again makes such a distinction uncertain, if confirmed this result would support hypothesis~\textbf{H3} that a typically unmovable object favors an egocentric mental model. As with translations, however, this effect appears to be very small. Unlike translations though, this effect would lead people toward the \emph{opposite} mapping compared to Scene~1. Yet, because of its small strength it does not seem to be sufficient to change the overall preference for allocentric rotations.

In contrast to the other scenes, Scene~3 showed a very marked preference for egocentric rotations ($r{=}0.57$, $\sigma_r{=}0.04$). This is again strong evidence for~\textbf{H1} that manipulating an object viewed from inside is preferably accomplished egocentrically.

Scene~4 showed the same pattern as Scenes~1 and~2, i.e.~we saw a preference for allocentric rotations ($r{=}0.28$, $\sigma_r{=}0.11$). Yet, egocentric rotations were still rated as natural by \mbox{little} more than half of the participants. Since the preferred mapping in Scene~1 was already allocentric, we cannot provide any particular support for~\textbf{H2}. If anything, the smaller effect size compared to Scene~1 tends to disprove~\textbf{H2} since the egocentric mapping was more readily accepted in this configuration.

\begin{figure}[t]
  \centering
  \includegraphics[width=\columnwidth]{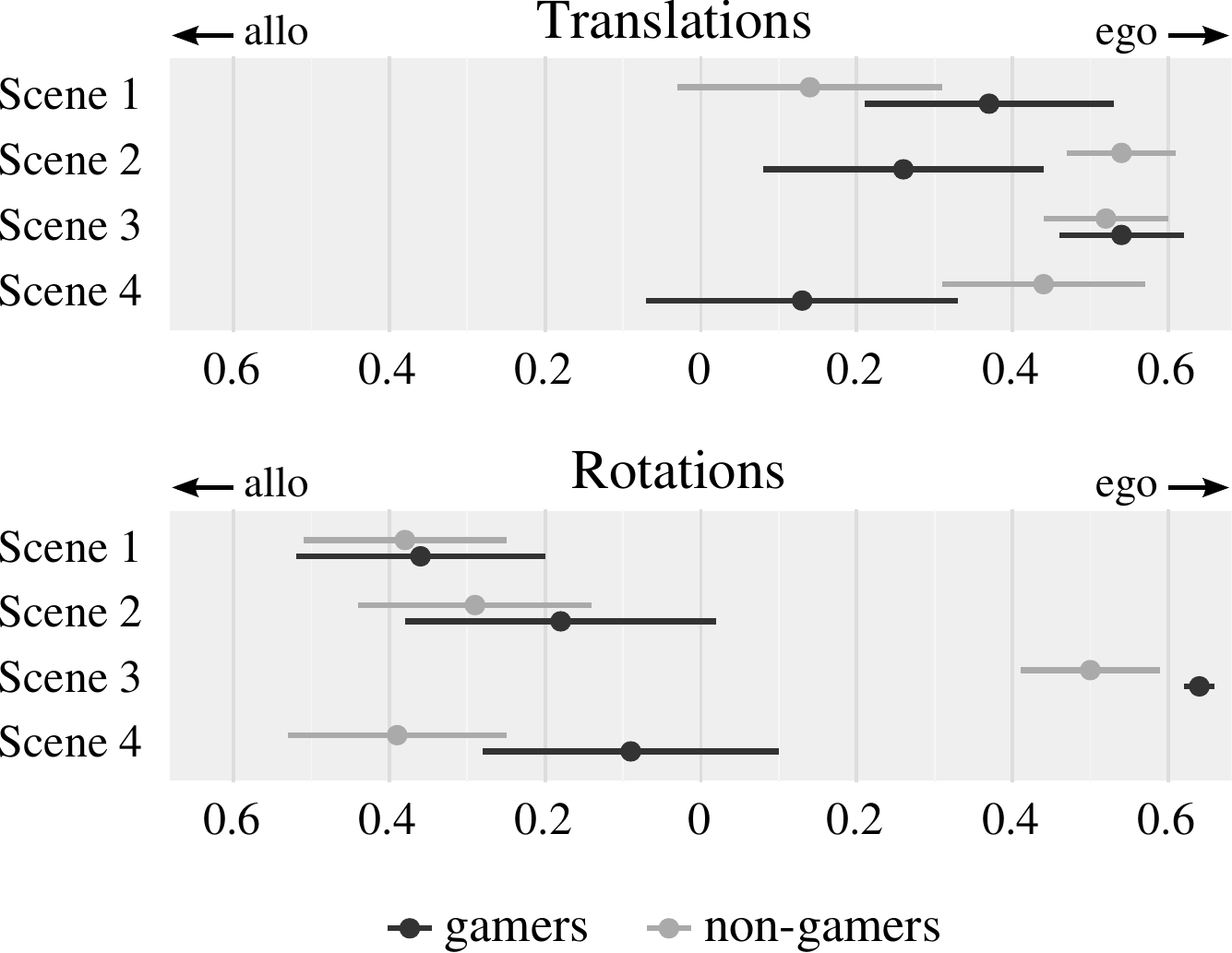}
  \caption{Effect sizes and standard error for the preferred mapping in the ``gamers'' and the ``non-gamers'' groups.}
  \label{fig:gamers}
\end{figure}

\subsection{Influence of gaming experience}

We conducted a second analysis of the translation and rotation results by splitting participants into two groups: those with regular experience with 3D~video games (\emph{gamers} group, 14~participants), and those who reported to seldom or never play such video games (\emph{non-gamers} group, 16 participants).
Although the preference for each mapping in each scene remained the same for both groups, in several cases the effect size was different---i.e.~the preference was less marked for one group than for the other (\autoref{fig:gamers}).

In Scene~1, non-gamers apparently had a weaker preference for egocentric translations ($r{=}0.14$, $\sigma_r{=}0.17$) compared to gamers ($r{=}0.37$, $\sigma_r{=}0.16$). Since Scene~1 can be considered as ``baseline,'' we can thus observe that---with some reservations due to the standard error---non-gamers may not actually have a strong \textit{a~priori} preference for translations, and that gaming experience may create a bias for the egocentric mapping. Concerning rotations, the preference was nearly identical between gamers ($r{=}0.36$, $\sigma_r{=}0.16$) and non-gamers ($r{=}0.38$, $\sigma_r{=}0.13$). It thus appears that gaming experience does not produce any such bias for rotations.

In Scene~2, the gamers' preference for egocentric translations was of a comparable level to Scene~1 ($r{=}0.26$, $\sigma_r{=}0.18$). The non-gamers' preference, however, was much more marked than in Scene~1 ($r{=}0.54$, $\sigma_r{=}0.07$). Gamers thus seem to be less affected by a change of scene contents than non-gamers, possibly because their experience makes them more tolerant to using either mapping in various virtual environments---albeit with a persistent bias toward the egocentric mapping. The difference between the two groups was less clear for rotations, but still hinted at a similar trend (gamers: $r{=}0.18$, $\sigma_r{=}0.20$; non-gamers: $r{=}0.29$, $\sigma_r{=}0.15$). These results may thus support hypothesis~\textbf{H3}, but only for non-gamers.

For Scene~3, the preference for egocentric translations was as strongly marked for gamers ($r{=}0.54$, $\sigma_r{=}0.08$) as it was for non-gamers ($r{=}0.52$, $\sigma_r{=}0.08$). We can thus infer that the conditions in Scene~3 (viewing the manipulated object from inside) had a strong enough effect to overcome the presumed tolerance of gamers for their non-preferred (allocentric) mapping. Rotations also showed a clear preference for the egocentric mapping among both groups (gamers: $r{=}0.64$, $\sigma_r{=}0.02$; non-gamers: $r{=}0.50$, $\sigma_r{=}0.09$), confirming hypothesis~\textbf{H1}.
\looseness=-1

In Scene~4, like in Scene~2, non-gamers had a strong preference for egocentric translations ($r{=}0.44$, $\sigma_r{=}0.13$), while gamers were more tolerant of either mapping ($r{=}0.13$, $\sigma_r{=}0.20$). We observed the same pattern with rotations: non-gamers had a preference for allocentric rotations ($r{=}0.39$, $\sigma_r{=}0.14$), while gamers were more neutral ($r{=}0.09$, $\sigma_r{=}0.19$). Although these results still contradict~\textbf{H2}, they are consistent with our above assumption that gamers may be more tolerant to using different mappings in various virtual environments.

\begin{figure*}[t]
  \includegraphics{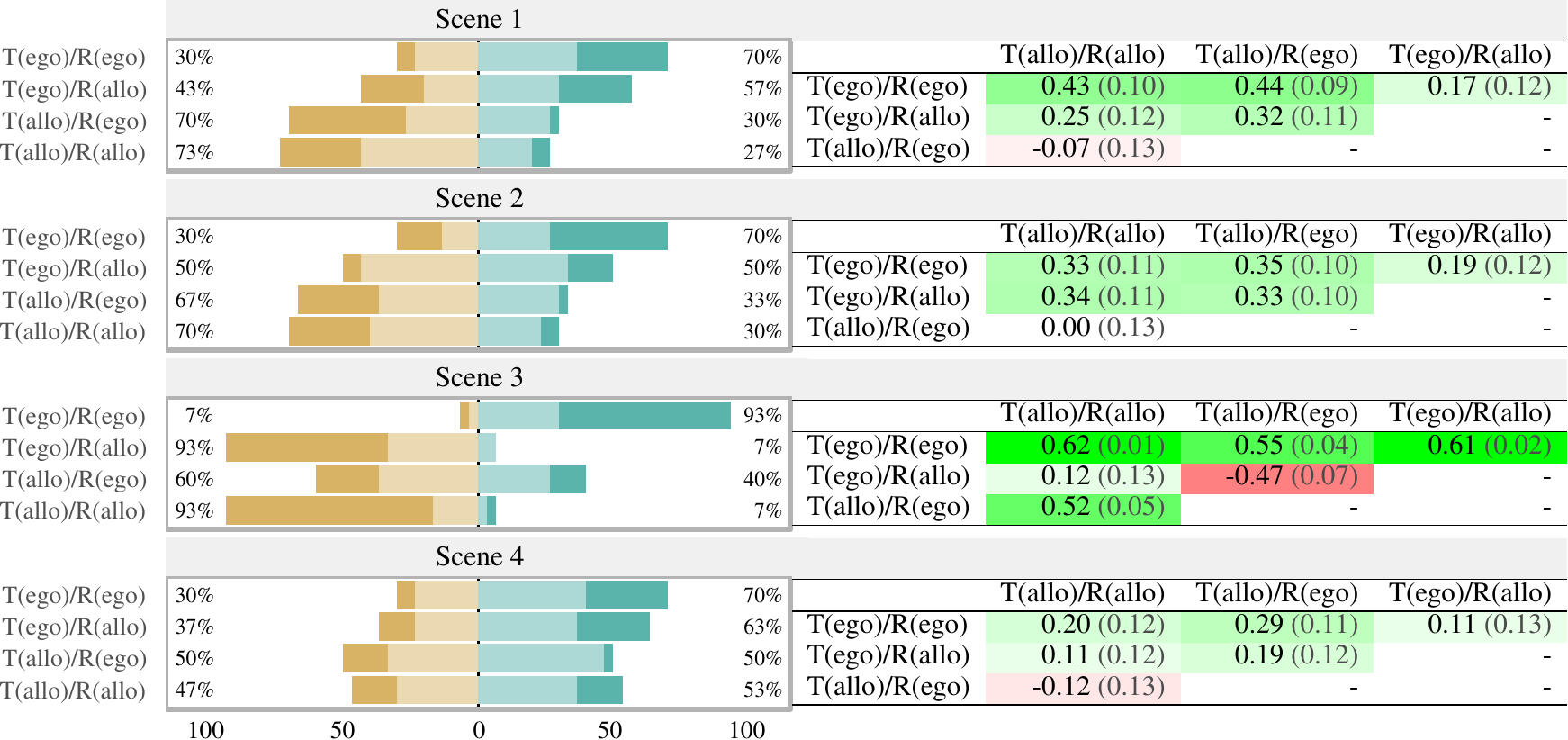}
  \caption{Ratings in the full 6-DOF~manipulation tasks (simultaneous translations/rotations), along with the pairwise effect sizes. The notation ``\tr{tmap}{rmap}'' refers to a translation mapping ``tmap'' and a rotation mapping ``rmap''.}
  \label{fig:part2}
\end{figure*}

\subsection{Simultaneous translations/rotations}

\autoref{fig:part2} shows the ratings for each combination of allocentric and egocentric translations and rotations, in tasks involving full 6-DOF manipulation.
For conciseness, we use the notation \tr{tmap}{rmap} in which ``tmap'' describes the translation mapping and ``rmap'' describes the rotation mapping.
The results for Scenes~1 and~2 both present a similar pattern: pairwise differences in ratings were comparatively larger between~T(ego) and~T(allo) combinations (medium effect~sizes) than between~R(ego) and~R(allo) combinations (small to zero effect~sizes). It thus seems that, in these two scenes, the choice of translation mapping is more important than the rotation mapping when performing both simultaneously. This appears to disprove hypothesis~\textbf{H4}.

As expected from the translation-only results, T(ego)~combinations were rated higher than T(allo)~combinations in all scenes except Scene~3. In these scenes, however, the~\tr{ego}{ego} mapping was apparently preferred to the~\tr{ego}{allo} mapping. This is surprising because our results for the first part of the experiment showed a preference for allocentric rotations in such scenes. Moreover, the~\tr{allo}{allo} mapping also seems to be preferred to the~\tr{allo}{ego} mapping in Scenes~1 and~4, with a tie in Scene~2. Although these differences are below the standard error, they nevertheless hint at a similar trend in each of these scenes. In addition to the dominance of the translation mapping, there might thus be a preference for having the same mapping for both translations and rotations.

Overall, the four combinations were given comparable ratings in Scenes~1 and~2, the differences between each combination were thus also comparable. These differences were, however, more uniform in Scene~2 than in Scene~1. This could be explained by the lower number of strongly negative ratings for the~\tr{ego}{allo} and~\tr{allo}{ego} combinations---i.e.~two combinations that featured an egocentric mapping. This is consistent with the previously identified small possible effect that would slightly reinforce the preference for the egocentric mapping in Scene~2. It also provides additional (if small) evidence for hypothesis~\textbf{H3}.

Scene~3 showed a strong difference between the fully egocentric (\tr{ego}{ego}) combination and the three other combinations. Almost all participants rated the former as natural, whereas the latter (not fully egocentric) three were largely rated as unnatural. Still, among these three lowest-rated combinations, \tr{allo}{ego}---the only one to feature egocentric rotations---was rated noticeably higher, despite consisting of two opposite mappings. We can hypothesize that this is due to the larger influence of rotations on visual flow when the manipulated object is viewed from inside. In such a situation, the positive effects of having rotations that match the preferred mapping (egocentric, as per our previous results) appear to noticeably alleviate the negative effects of a non-preferred (allocentric) translation mapping, even though the translation and rotation mappings are different. In any case, these results confirm again the importance of an egocentric mapping when the manipulated object is viewed from inside (hypothesis \textbf{H1}).

\begin{figure*}[t]
  \centering
  \includegraphics{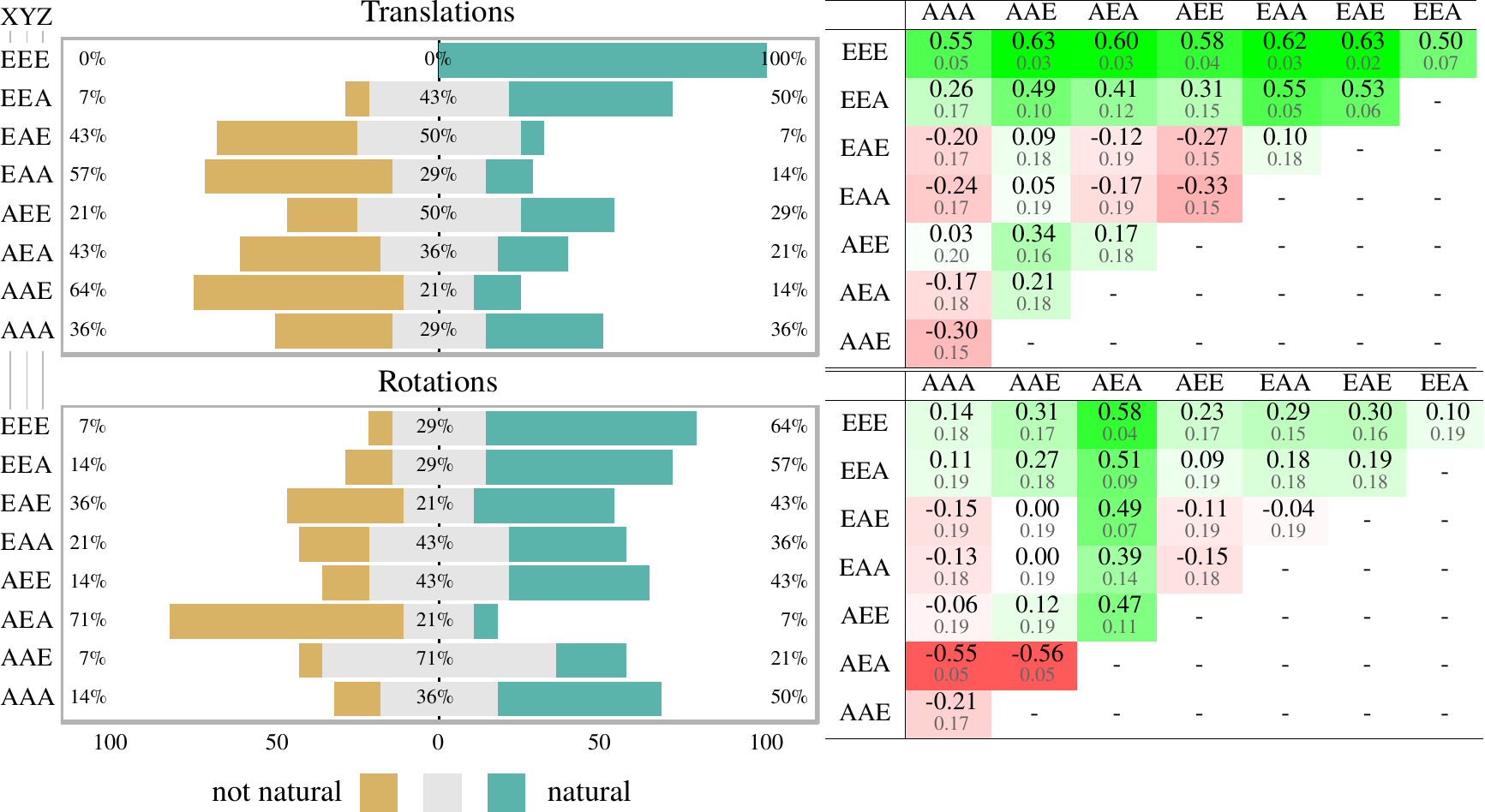}
  \caption{Ratings and pairwise effect sizes in the per-axis inversion part (labels are in the form ``XYZ'': A=allocentric, E=egocentric).}
  \label{fig:part3}
\end{figure*}

\subsection{Per-axis inversion}

\autoref{fig:part3} shows the ratings given by participants in the third part of the experiment, in which the direction of motion along each manipulation axis was alternatively inverted. The $x$-axis ran along the left-right direction of the tablet's screen (which was itself held in landscape orientation), the $y$-axis was aligned with the top-down direction of the screen, and the $z$-axis was orthogonal to the screen plane. Unlike during the previous parts, we asked for ratings on a 3-point Likert~scale.

For translations, one configuration was clearly preferred over the others: the fully egocentric mapping. This is consistent with the results of the previous parts that revealed a preference for egocentric translations in Scene~1. Given the strength of this effect (100\%~of participants rated this mapping as natural, far above the other configurations) there does not seem to be any advantage to gain from a mixed mapping for~translations.

However, one question that still remains is whether some manipulation axes have more influence than others in the perceived naturalness of a mapping (hypothesis~\textbf{H5}). One such pattern seems to be visible in the above results: configurations where the $y$-axis was inverted (i.e.~egocentric) were consistently rated higher than the others---with the exception of the fully allocentric configuration. We thus conducted a further analysis by merging the ratings in egocentric and allocentric groups depending on the state (inverted or not) of each manipulation axis. The results indeed revealed a larger effect of the $y$-axis ($r{=}0.35$, $\sigma_r{=}0.07$), compared to the effects of the $x$-axis ($r{=}0.18$, $\sigma_r{=}0.09$) and the $z$-axis ($r{=}0.05$, $\sigma_r{=}0.10$). The reason why the fully allocentric was comparatively rated higher, despite its non-inverted $y$-axis, could be explained by a preference for a consistent mapping between all axes---similar to the possible preference for a consistent mapping between translations and rotations discussed in the previous section. Nevertheless, the larger influence of the $y$-axis appears to validate our \textbf{H5}~hypothesis for translations.

For rotations, no configuration was unanimously favored and the differences between each configuration were smaller overall than for translations. Merging the ratings into groups according to each manipulation axis confirmed that the influences of all axes were limited ($x$:~$r{=}0.17$, $\sigma_r{=}0.09$; $y$:~$r{=}0.00$, $\sigma_r{=}0.09$; and $z$:~$r{=}0.11$, $\sigma_r{=}0.09$), which does not confirm \textbf{H5} for rotations. Yet, one configuration seems to have been strongly disliked compared to all the others: the allo-ego-allo~(AEA) mapping. This result is unexpected, and we cannot find any reason that could explain such a sharp drop in ratings. This aspect would thus require further \mbox{investigation}.

\section{Conclusion}

The ``baseline'' preference, i.e.~which alternative was found most natural with no prior exposure to the interface and minimal contextual cues, was \mbox{\emph{egocentric translations}} and \mbox{\emph{allocentric rotations}} among all users. In the gamers group, the baseline for translations was less marked than in the non-gamers group. However, the baseline for rotations was similar.

Regarding contextual cues, \textbf{H1} was strongly supported by our results: when a manipulated object is viewed from inside, the mapping should be egocentric for both translations and rotations. Surprisingly, \textbf{H2} was not supported. It appears that manipulating an object within a fixed virtual environment does not induce a preference for an allocentric mapping. There was limited support for \textbf{H3} that a typically static object should be manipulated with an egocentric mapping. The results hint at a possible weak effect, though not sufficient to overcome other factors (such as the baseline preference).

Gamers seem to be less influenced by the scene contents than non-gamers---except when the manipulated object was viewed from inside. This could mean that gamers are more tolerant to encountering various mappings in 3D~applications.
Yet, the overall preferred mappings were still the same in both groups. We recommend, therefore, that translations-only mappings should be made egocentric in all cases, and that rotation-only mappings should be made allocentric in all cases \emph{except} when the manipulated object is viewed from inside---in which case it should be made egocentric.

In full 6-DOF~mappings, where users perform both translations and rotations simultaneously, the choice of a ``good'' translation mapping (egocentric, according to the results of part~1) seems to be more important than the choice of rotation mapping, thus disproving the null hypothesis~\textbf{H4}. However, when the manipulated object is viewed from inside, \emph{both} mappings should be egocentric. In addition, there seems to be a positive effect of having the same mapping for translations and rotations. Therefore, we recommend that a 6-DOF~manipulation mapping should be T(ego)/R(ego) (i.e.~fully egocentric) in~all~cases.

We saw no benefit of a ``mixed'' mapping that selectively inverts some of the manipulation axes, compared to a ``fully'' egocentric or allocentric mapping. Yet, if such a mixed mapping must be implemented, the $y$-axis (along the vertical direction of the device's screen) appears to have more influence on the perceived naturalness of translations. This would support hypothesis~\textbf{H5}, but only for translations. For reasons we have yet to explain, one mixed mapping for rotations (allo-ego-allo) was found much less natural than all the~others.


\bibliographystyle{abbrv}
\balance
\bibliography{article}

\begin{thebibliography}{10}

\bibitem{bartlett00}
J.~Bartlett.
\newblock {R}ock 'n' {S}croll is here to stay.
\newblock {\em {IEEE} Computer Graphics and Applications}, 20(3):40--45, May
  2000.

\bibitem{bowman04}
D.~A. Bowman, E.~Kruijff, J.~J. LaViola, Jr., and I.~Poupyrev.
\newblock {\em {3D} User Interfaces: Theory and Practice}.
\newblock Addison-Wesley, Boston, 2004.

\bibitem{burgess04}
N.~Burgess, H.~J. Spiers, and E.~Paleologou.
\newblock Orientational manoeuvres in the dark: Dissociating allocentric and
  egocentric influences on spatial memory.
\newblock {\em Cognition}, 94(2):149–166, 2004.

\bibitem{chan03}
A.~H.~S. Chan, V.~W.~Y. Shum, H.~W. Law, and I.~K. Hui.
\newblock Precise effects of control position, indicator type, and scale side
  on human performance.
\newblock {\em The International Journal of Advanced Manufacturing Technology},
  22(5--6):380--386, 2003.

\bibitem{diaz05}
D.~D. Diaz and V.~K. Sims.
\newblock Accidental inversion during three-dimensional orientational control.
\newblock {\em Proceedings of the Human Factors and Ergonomics Society Annual
  Meeting}, 49(13):1248--1250, 2005.

\bibitem{fitts51}
P.~M. Fitts.
\newblock Engineering psychology and equipment design.
\newblock In S.~S. Stevens, editor, {\em Handbook of experimental psychology},
  pages 1287--1340. Wiley, 1951.

\bibitem{fitts53}
P.~M. Fitts and C.~M. Seeger.
\newblock {SR} compatibility: Spatial characteristics of stimulus and response
  codes.
\newblock {\em Journal of Experimental Psychology}, 46(3):199--210, 1953.

\bibitem{fritz12}
C.~O. Fritz, P.~E. Morris, and J.~J. Richler.
\newblock Effect size estimates: Current use, calculations, and interpretation.
\newblock {\em Journal of Experimental Psychology: General}, 141(1):2--18,
  2012.

\bibitem{hinckley11}
K.~Hinckley and H.~Song.
\newblock Sensor synaesthesia: Touch in motion, and motion in touch.
\newblock In {\em Proc.\ CHI}, pages 801--810, New York, 2011. ACM.

\bibitem{issartel16}
P.~Issartel, F.~Gu{\'e}niat, T.~Isenberg, and M.~Ammi.
\newblock Analysis of locally coupled 3d manipulation mappings based on mobile
  device motion.
\newblock arXiv preprint 1603.07462, Mar. 2016.

\bibitem{kaminaka85}
M.~S. Kaminaka and E.~A. Egli.
\newblock Lever controls on specialised farm equipment: Some control/response
  stereotypes.
\newblock {\em Applied Ergonomics}, 16(3):193--199, 1985.

\bibitem{klatzky98}
R.~L. Klatzky.
\newblock Allocentric and egocentric spatial representations: Definitions,
  distinctions, and interconnections.
\newblock In {\em Spatial Cognition}, pages 1--17. Springer,
  Berlin\discretionary{/}{}{/}Heidelberg, 1998.

\bibitem{neale13}
S.~Neale, W.~Chinthammit, C.~Lueg, and P.~Nixon.
\newblock {RelicPad}: A hands-on, mobile approach to collaborative exploration
  of virtual museum artifacts.
\newblock In {\em Proc.\ INTERACT}, pages 86--103,
  Berlin\discretionary{/}{}{/}Heidelberg, 2013. Springer.

\bibitem{poupyrev98}
I.~Poupyrev, T.~Ichikawa, S.~Weghorst, and M.~Billinghurst.
\newblock Egocentric object manipulation in virtual environments: Empirical
  evaluation of interaction techniques.
\newblock {\em Computer Graphics Forum}, 17(3):41--52, 1998.

\bibitem{rekimoto96}
J.~Rekimoto.
\newblock Tilting operations for small screen interfaces.
\newblock In {\em Proc.\ UIST}, pages 167--168, New York, 1996. ACM.

\bibitem{warrick47}
M.~J. Warrick.
\newblock Direction of movement in the use of control knobs to position visual
  indicators.
\newblock {\em Psychological Research on Equipment Design}, pages 137--146,
  1947.

\bibitem{weberg01}
L.~Weberg, T.~Brange, and {\AA}.~\mbox{Wendelbo-} \mbox{Hansson}.
\newblock A piece of butter on the {PDA} display.
\newblock In {\em CHI Extended Abstracts}, pages 435--436, New York, 2001. ACM.

\bibitem{wiebe09}
J.~Wiebe and K.-P.~L. Vu.
\newblock Application of population stereotypes to computerized tasks.
\newblock In {\em Proc.\ Human Interface and the Management of Information},
  pages 718--725. Springer, Berlin\discretionary{/}{}{/}Heidelberg, 2009.

\end{thebibliography}
\end{document}